\documentstyle[prd,aps,preprint]{revtex}

\newcommand\be{\begin{equation}}
\newcommand\ee{\end{equation}}
\newcommand\bea{\begin{eqnarray}}
\newcommand\eea{\end{eqnarray}}

\newcommand\lt{\left}
\newcommand\rt{\right}

\newcommand\eq[1]{Eq.~(\ref{#1})}
\newcommand\eqs[2]{Eqs.~(\ref{#1}) and (\ref{#2})}

\newcommand\mpl{M_{\rm Pl}}

\begin{document}
\preprint{LANCS-TH/9613, RESCEU-20/96, hep-ph/9606412}
\draft
\tighten

\title{More varieties of hybrid inflation}
\author{David H. Lyth}
\address{School of Physics and Chemistry, Lancaster University, 
Lancaster, LA1 4YB, U.K.}
\author{Ewan D. Stewart}
\address{Research Center for the Early Universe, School of Science,
University of Tokyo, Bunkyo-ku, Tokyo 113, Japan}
\date{June 1996}
\maketitle
\begin{abstract}
It is pointed out that hybrid inflation can be implemented with the
inflaton field rolling away from the origin instead of towards it.
This `inverted' hybrid inflation has a spectral index $ n < 1 $,
in contrast with ordinary hybrid inflation which has $ n \gtrsim 1 $,
so a measured value of $n$ substantially different from 1 would
distinguish the two.
Other generalisations of hybrid inflation are also considered.
\end{abstract}

\section{Introduction}

The most attractive models of inflation at present are the
`hybrid' inflation models \cite{l90,AdamsFreese,LIN2SC},
in which inflation ends due to the interaction of the inflaton field
with other fields.
They share with the very first models \cite{new} the virtue
that the inflaton field variation is small on the Planck scale
\cite{CLLSW}, without requiring their fine-tuning.
They have been widely studied, both in the original versions
\cite{l90,AdamsFreese,LIN2SC,LL2,LIN2SC2,CLLSW,silvia}
and in modified ones 
\cite{LIN2SC2,ewansgrav,qaisar,mutated,lazpan,dave,lisa,glw}. 

Inflation generates an adiabatic density perturbation and
gravitational waves.
The former is supposed to be the origin of 
large scale structure and, together with a possible contribution from
the gravitational waves, of the cosmic microwave background
(cmb) anisotropy. 
The spectrum of the density perturbation is 
conveniently specified by a quantity $\delta_H^2$, whose scale 
dependence is $\delta_H^2\propto k^{n-1}$ where $n$ is the spectral 
index. The spectrum of the gravitational waves
is conveniently specified by their relative contribution $R$
to the mean-square low multipoles of the cmb anisotropy seen by a randomly 
placed observer.

Within the usual paradigm of a single slow-rolling 
field\footnote
{A more general slow-roll paradigm \cite{misao} leaves the
gravitational waves unchanged but increases $\delta_H$, so that
\eq{dh} becomes a lower bound, \eq{R} an upper bound and \eq{n} is no
longer valid.}
(which includes most versions of hybrid inflation considered up to
now, and we shall also restrict ourselves to this simple case) the
predictions for $\delta_H^2$ \cite{adpred,lyth85}, $n$
\cite{LL1,salopek} and $R$ \cite{starobinsky} are
\bea
\label{dh}
\delta_H^2(k) &=& \frac1{75\pi^2} \frac{V^3}{V'^2} \\
\label{n}
n-1 &=& - 3 \lt( \frac{V'}{V} \rt)^2 + 2 \frac{V''}{V} \\
\label{R}
R &=& 6 \lt( \frac{V'}{V} \rt)^2
\eea
The units are $ \hbar = c = \mpl \equiv (8\pi G)^{-1/2} = 1 $,
and the potential $V$ and its derivatives are to be evaluated
when cosmological scales leave the horizon $N$ $e$-folds
before the end of inflation, where $N=50$ to $25$ and is given by
$N=\int (V/V')d\phi$. 
In models with a small field variation, including in particular
hybrid inflation models, $R$ is unobservably small \cite{davidnew}
and $n$ is either very close to $1$ or is given by $ n-1 = 2 V''/V $. 

At present observation is consistent with $R=0$, and with this value
the COBE measurement of the low multipoles of the cmb anisotropy gives
\cite{delh} $\delta_H=1.94\times 10^{-5}$.
The present constraint on $n$ is only \cite{constraint}
$0.7<n<1.4$,
though eventually one can expect a measurement with an accuracy
$\Delta n\sim .01$ \cite{davidprep}.

In most versions of hybrid inflation proposed so far,
the inflaton field is near a minimum of the potential.
For reasonable choices of the parameters the model can give
practically any value of $n$ in the observed range {\em bigger} than
1.
Two versions \cite{mutated,lazpan} have a mildly concave potential
leading to $n=0.93$ to $0.97$. 
We are going to point out the existence of `inverted' hybrid
inflation, which works near a maximum of the potential.
As a result it can give practically any value {\rm below} 1.

The outline of the paper is as follows.
In the next section we study inverted hybrid inflation,
and in the following one mention some generalizations of it.
In Section~\ref{gmhi} we consider a whole class of models that
generalize the mutated hybrid inflation of Ref.~\cite{mutated}.
In Section~\ref{con} we give our conclusions.

\section{Inverted hybrid inflation}
\label{ihi}

In the usual models of hybrid inflation $\phi$ is rolling towards
zero. The potential is typically of the form
\be
V = V_0 + \frac12 m^2 \phi^2 + \ldots
\ee
and is dominated by the term $V_0$. This term arises because some
other field $\psi$ is held at the origin by its interaction with
$\phi$. When $\phi$ falls below some critical value $\phi_{\rm c}$,
the other field rolls to its vacuum value so that $V_0$
disappears and inflation ends. This model gives negligible $R$ and
$n-1= 2m^2/V_0$.

We consider instead the opposite case of `inverted' hybrid inflation,
where $\phi$ rolls away from the origin and
\begin{eqnarray}
V & = & V_0 - \frac12 m^2 \phi^2 + \ldots \\
& = &
V_0 \left[ 1 - \frac12 \left(\frac{\phi}{f}\right)^2 + \ldots \right]
\label{pot}
\end{eqnarray}
We again assume that $V_0$ dominates, and have introduced the
parameter $f^2=V_0/m^2$.
This gives 
$ n-1 = - 2 m^2 / V_0 = - 2 / f^2 $.
The present observational lower bound $n>0.7$ 
\cite{constraint}
requires $f>2.6$ and any
future tightening will increase the required value of $f$. 

In this model, gravitational waves are negligible and
the  normalization is
\be
V_0 = 7 \times 10^{-8} (1-n)^2 \phi_{\rm c}^2 e^{-(1-n)N}
\ee
where we have used the results of Section~\ref{gmhi}.

A complete model should specify the mechanism which ends inflation.
In ordinary hybrid inflation one considers a potential
\be
V = V_0 + \frac{1}{2} m_\phi^2 \phi^2 - \frac{1}{2} m_\psi^2 \psi^2
+ \frac{1}{2} \lambda \phi^2 \psi^2 + \ldots
\label{full}
\ee
with $ m_\phi^2 \ll V_0 \lesssim m_\psi^2 $.
One possiblity for inverted hybrid inflation is just to reverse the
signs of all the terms
\be
V = V_0 - \frac{1}{2} m_\phi^2 \phi^2 + \frac{1}{2} m_\psi^2 \psi^2
- \frac{1}{2} \lambda \phi^2 \psi^2 + \ldots
\label{first}
\ee
Then $\psi$ will be constrained to zero for
\be
\phi < \phi_{\rm c} = \frac{m_\psi}{\sqrt{\lambda}}
\ee
and will roll away from zero when $\phi$ becomes larger than
$\phi_{\rm c}$.
This model could be realized in supersymmetry using the superpotential
\be
W = \lt( \Lambda^2 + \frac{\lambda\Phi^2\Psi^2}{\Lambda^2} \rt) \Xi
\ee
The corresponding globally supersymmetric scalar potential is
\be
V = \lt| \Lambda^2 + \frac{\lambda\Phi^2\Psi^2}{\Lambda^2} \rt|^2
+ \frac{4\lambda^2}{\Lambda^4}
\lt( |\Phi|^2 + |\Psi|^2 \rt) |\Phi|^2 |\Psi|^2 |\Xi|^2
\ee
Writing $ |\Phi| = \phi / \sqrt{2} $ and $ |\Psi| = \psi / \sqrt{2} $,
minimising with respect to $\arg\Phi^2\Psi^2$, and 
assuming $\Xi$ is held at zero, gives
\be
V = \lt( \Lambda^2 - \frac{\lambda\phi^2\psi^2}{4\Lambda^2} \rt)^2
= \Lambda^4 - \frac{1}{2} \lambda \phi^2 \psi^2 + \ldots
\ee
Adding soft supersymmetry breaking masses\footnote{
Note that one would in general expect
$ m_\phi \sim m_\psi \sim \Lambda^2 $ but as we have
$ V = |W_\Xi|^2 $ and $ W_\Phi = W = \Xi = 0 $ we can use the
method of Ref.~\cite{ewansgrav} to allow $ m_\phi \ll \Lambda^2 $.}
of the required sign gives
\be
V = \Lambda^4 - \frac{1}{2} m_\phi^2 \phi^2
+ \frac{1}{2} m_\psi^2 \psi^2
- \frac{1}{2} \lambda \phi^2 \psi^2 + \ldots
\ee

An alternative way of implementing inverted hybrid inflation is given
by
\be
\label{alt}
V = V_0
- \frac{1}{2} m_\phi^2 \phi^2
- \frac{1}{2} m_\psi^2 \psi^2
- \frac{1}{2} m_\chi^2 \chi^2
+ \frac{1}{2} \lambda_\phi^2 \phi^2 \chi^2
+ \frac{1}{2} \lambda_\psi^2 \psi^2 \chi^2
+ \frac{1}{4} \lambda_\chi^2 \chi^4
\ee
with $ m_\phi^2 \ll V_0 \lesssim m_\psi^2 , m_\chi^2 $.
Here $\psi$ will be constrained to zero if
$ \chi > m_\psi / \lambda_\psi $.
The minimum of $\chi$'s potential is at
\be
\chi = \frac{ \sqrt{ m_\chi^2 - \lambda_\phi^2 \phi^2 } }
{ \lambda_\chi }
\ee
assuming $ \psi = 0 $ and $ \phi < m_\chi / \lambda_\phi $.
Therefore for
\be
\phi < \phi_{\rm c}
= \frac{ \sqrt{ \lambda_\psi^2 m_\chi^2 - \lambda_\chi^2 m_\psi^2 } }
{ \lambda_\phi \lambda_\psi }
\ee
$\psi$ will be constrained to zero.
Clearly we require $ \lambda_\psi m_\chi > \lambda_\chi m_\psi $.
Also, to ensure that the terms involving $\chi$ in \eq{alt} make a
negligible contribution to the effective potential during inflation,
we require $ \lambda_\phi m_\chi \ll \lambda_\chi m_\phi $.
The contribution of the $\phi$ dependence of $\chi$ to the effective
kinetic terms during inflation can be neglected if
$ \lambda_\phi \lambda_\psi m_\chi \ll \lambda_\chi^2 m_\psi $.

Previous authors 
\cite{nontherm,bingall,natural,natural2,paul}
considered the potential (\ref{pot})
in the context of a single-field model,
and assumed that it holds until $\phi\simeq f$,
after which $\phi$ settles down to the minimum located at a value
somewhat bigger than $f$.
Since observation requires $f\gg 1$ (the Planck scale in our units),
this places the model outside the regime of ordinary
particle theory so that one can hardly justify the form of the
potential.
It has been suggested that $\phi$ might be identified with one of the
superstring moduli \cite{bingall,natural2,paul}, but attempts 
\cite{bingall,natural2,macorra,bento} to construct a specific model 
using this idea have not been successful.\footnote
{Ref.~\cite{bento} claims to have been successful but an analytic
calculation shows that their Eq.~7 gives a potential
$ V(\phi) \simeq - 4.3 - 0.01 \cos (12\phi) $ for the canonically
normalised field proportional to Im$S$ after minimizing with respect
to Re$S$.} 

We have avoided these difficulties by ending inflation 
earlier using the hybrid inflation mechanism. An alternative
would be to keep the single field $\phi$, but to 
suppose that its potential steepens soon after 
cosmological scales leave the horizon 
so that inflation again ends at some value $\phi_c\ll
f$. Such a proposal is not unreasonable, though it does 
postulate a lot of structure in the single potential $V(\phi)$.

\section{Generalizations of inverted hybrid inflation}

Instead of assuming that the inflationary potential is quadratic,
one can consider the possibility that it is of higher order. This might 
be because the quadratic term is absent, or it might be because 
one is not close to the origin though in that case there is no reason to 
suppose that a single term dominates. If the quadratic term is absent
one might have
\be
V = V_0 - \frac14 \lambda \phi^4 + \ldots
\label{quartic}
\ee
This form could arise from one of the generalizations of mutated hybrid 
inflation considered in the next section, and we use the results
derived there.
For $ \phi_{\rm c}^2 \gg V_0/(2N\lambda) $ the predictions are
independent of $\phi_{\rm c}$ and the same as for the non-hybrid case;
$ 1-n = 3/N $ with negligible gravitational waves and the
normalization of $\delta_H$ requiring $ \lambda \sim 10^{-12} $
independently of $V_0$.
In the opposite case, $1-n$ is reduced by a factor
$\phi_{\rm c}^2/[V_0/(2N\lambda)]$ and the value of $\lambda$ is reduced by
this factor cubed.

Including both a quadratic and a quartic term one could have 
\be
V = V_0 - \frac12 m^2 \phi^2 + \frac14\lambda \phi^4 
\ee
If inflation (after the observable Universe leaves the horizon)
takes place near the maximum it reduces to the quadratic inverted
hybrid inflation that we started out with. 
If it takes place near the minimum it reduces to the usual version of
hybrid inflation, and in the intermediate case one gets something
different. 

Finally, we should point out that a related model of inflation, 
 was proposed a long time ago \cite{early}.
In it, the inflaton is rolling away from the origin and it triggers the
GUT Higgs transition at or before the end of inflation.
The potential for this model is very complicated
and it is not clear whether the Higgs potential is supposed to dominate the 
energy density. If there is a regime of parameter space in which it does,
then the model would represent the first version of inverted hybrid 
inflation (and in fact the first version of hybrid inflation of any 
kind).

\section{Generalizations of mutated hybrid inflation}
\label{gmhi}

Here we consider potentials of the form
\be
\label{Vm}
V = V_0 - \frac{\sigma}{p} \psi^p + \frac{\lambda}{q} \psi^q \phi^r
\ee
with $ p \neq q $.
The opposite case of $p=q$ simply gives an ordinary or inverted
hybrid inflation model where $\psi$ is held at zero during inflation.
In mutated hybrid inflation \cite{mutated}, and its generalisations
\cite{lazpan} considered here, $\psi$ is held close to zero, but not
at zero, during inflation.
An effective potential for the inflaton $\simeq\phi$ is then generated
from the couplings without requiring any additional term and, as we
shall see, it can take unusual forms.
In the case that this contribution to the effective potential is not
the dominant one, it will not determine the spectrum but may still
determine when inflation ends.

We assume $\psi>0$ and $\phi>0$.
Then to get a model which, for fixed $\phi$, has a minimum at
$\psi=\psi_*$ with $ \psi_* > 0 $, we require $ \sigma \lambda > 0 $
and $ (q-p) \sigma > 0 $. See \eqs{mvp}{mvpp} respectively below.
$ \sigma > 0 $ corresponds to a generalisation of mutated hybrid
inflation with the inflaton $\phi$ rolling towards zero, while
$ \sigma < 0 $ corresponds to a general inverted mutated hybrid
inflation model with $\phi$ rolling away from zero.

Now
\be
\label{mvp}
V_\psi = - \sigma \psi^{p-1} + \lambda \psi^{q-1} \phi^r 
\ee
and so $ V_\psi = 0 $ when $ \psi = 0 $ (for $p\geq2$ and $q\geq2$) or
\be
\psi = \psi_*
\equiv \left( \frac{\sigma}{\lambda} \right)^{\frac{1}{q-p}}
\phi^{-\frac{r}{q-p}}
\ee
Now
\be
\label{mvpp}
\lt.V_{\psi\psi}\rt|_{\psi=\psi_*} = (q-p) \sigma \psi_*^{p-2}
\ee
For simplicity we will assume that
$ V_{\psi\psi}|_{\psi=\psi_*} \gg V_0 $ so that $\psi$ is held firmly
at $\psi=\psi_*$ during inflation.
Then
\be
V|_{\psi=\psi_*} = V_0 - \lt( \frac{q-p}{pq} \rt) \sigma \psi_*^p
\ee
and the kinetic terms evaluated along $\psi=\psi_*$ are
\be
\frac{1}{2}
\lt[ 1 + \lt( \frac{r}{q-p} \rt)^2 \frac{\psi_*^2}{\phi^2} \rt]
\lt( \partial \phi \rt)^2
\ee
Assuming $ \sigma \psi_*^p \ll V_0 $ so that $V_0$ dominates the
energy density and $ \psi_* \ll \phi $ so that the kinetic terms are
approximately canonical, we get the effective potential during
inflation
\be
V(\phi) = V_0 \left( 1 - \mu \phi^{-\alpha} \right)
\ee
where
\be
\mu = \lt( \frac{q-p}{pq} \rt)
\frac{ \sigma^{\frac{q}{q-p}} \lambda^{-\frac{p}{q-p}} }{ V_0 }
> 0
\ee
and
\be
\alpha = \frac{pr}{q-p} \neq 0
\ee
Now
\be
\frac{V'}{V} = \alpha \mu \phi^{-\alpha-1}
\ee
and
\be
\frac{V''}{V} = -\alpha (\alpha+1) \mu \phi^{-\alpha-2}
\ee
The number of $e$-folds to the end of inflation is given by
\bea
N = \int \frac{V}{V'} \,d\phi
& = & \frac{\phi^{\alpha+2}}{\alpha(\alpha+2)\mu}
\hspace{12em} \alpha > 0 \;\;{\rm or}\;\; \alpha < -2 \\
& = & \frac{1}{ |\alpha| \lt( 2 - |\alpha| \rt) \mu }
\lt( \phi_{\rm c}^{2-|\alpha|} - \phi^{2-|\alpha|} \rt)
\hspace{4em} -2 < \alpha < 0 \\
& = & - \frac{1}{2\mu} \ln \frac{\phi}{\phi_{\rm c}}
\hspace{15em} \alpha = -2
\eea
The COBE normalisation \cite{constraint} gives
\bea
5.3 \times 10^{-4} = \frac{V^{3/2}}{V'}
= \frac{ V_0^{1/2} \phi^{\alpha+1} }{ \alpha \mu }
& = & |\alpha|^{-\frac{1}{\alpha+2}}
\lt| \alpha + 2 \rt|^{\frac{\alpha+1}{\alpha+2}}
N^{\frac{\alpha+1}{\alpha+2}}
\mu^{-\frac{1}{\alpha+2}} V_0^{1/2}
\hspace{4em} \alpha > 0 \;\;{\rm or}\;\; \alpha < -2 \\
& = & \frac{ V_0^{1/2} }{ |\alpha| \mu }
\lt[ \phi_{\rm c}^{2-|\alpha|} - |\alpha| \lt( 2 - |\alpha| \rt) \mu N
\rt]^{ - \frac{|\alpha|-1}{2-|\alpha|} } 
\hspace{4em} -2 < \alpha < 0 \\
& = & \frac{ e^{2\mu N} V_0^{1/2} }{ 2 \mu \phi_{\rm c} }
\hspace{18em} \alpha = -2
\eea
The spectral index is given by
\bea
n \simeq 1 + 2 \frac{V''}{V}
& = & 1 - 2 \lt( \frac{\alpha+1}{\alpha+2} \rt) \frac{1}{N}
\hspace{8em} \alpha > 0 \;\;{\rm or}\;\; \alpha < -2 \\
& = & 1 - \frac{ 2 |\alpha| \lt( |\alpha| - 1 \rt) \mu }
{ \phi_{\rm c}^{2-|\alpha|} - |\alpha| \lt( 2 - |\alpha| \rt) \mu N }
\hspace{4em} -2 < \alpha < 0 \\
& = & 1 - 4\mu
\hspace{15em} \alpha = -2
\eea

Note that $ n < 1 $ in all cases except $ -1 \leq \alpha < 0 $.
However, to get $\alpha$ in this range would require either $ r < 1 $
or at least one of $p$, $q$ or $r$ to be negative.
The case $\alpha=-2$ is the one considered already in
Section~\ref{ihi}, and for this as well as other $\alpha$ in the range
$ -2 \leq \alpha < 0 $ we needed to specify that inflation ends at 
$ \phi = \phi_{\rm c} $. 
If inflation ends because $V_0$ stops dominating the energy density,
we will have $ \phi_{\rm c} \sim \mu^{1/\alpha} $, but 
higher order terms neglected in \eq{Vm} may end inflation
before this which would permit the desirable 
$ \phi_{\rm c} \lesssim 1 $. 

\subsection*{Supersymmetric implementations of generalized mutated
hybrid inflation}

Potentials of the form (\ref{Vm}) 
can be straightforwardly derived from supersymmetry along
the lines of Ref.~\cite{mutated} for the case $\sigma>0$
(corresponding to $\alpha>0$) and as in 
Section~\ref{ihi} for the opposite case.
In both cases, the superpotentials involved will be 
compatible with the method of
Ref.~\cite{ewansgrav} for avoiding fatal supergravity corrections.

In supersymmetry one would prefer $q$ and $r$ to be even if
$\sigma>0$, and $p$ to be even if $\sigma<0$.
Particularly natural possibilities are
$ \sigma \sim V_0 $ and $p=1$ or $2$,
and $ \sigma \sim - V_0 $ and $p=2$, with 
the $p=1$ and $p=2$ cases corresponding respectively
to a generic soft supersymmetry breaking
term for a singlet and a non-singlet.

The three simplest possiblities compatible with these ideas are
\begin{enumerate}
\item $p=1$, $q=2$, $r=2$, leading to $\alpha=2$. This is the original
mutated hybrid inflation model \cite{mutated}.
\item $p=2$, $q=1$, $r=1$, leading to $\alpha=-2$ which is the 
inverted hybrid inflation model of 
Section II.
\item $p=2$, $q=1$, $r=2$, leading to $\alpha=-4$ which is the quartic
inverted hybrid inflation model of Section III.
\end{enumerate}.

\section{Conclusion}
\label{con}

The most important new model that we have discussed is inverted hybrid
inflation.
In contrast with all other known hybrid inflation models it can give a
spectral index {\em significantly} below 1.
The other models we have discussed either reproduce already known
possibilities (though with a different prescription for the field
value at which inflation ends), or else add to the list of models 
which give a spectral index {\em slightly} below 1.
A future measurement of $n$ will be a powerful discriminator between
hybrid inflation models.

\subsection*{Acknowledgements}

DHL acknowledges support from PPARC, and from the European Commission
under the Human Capital and Mobility programme, contract
No.~CHRX-CT94-0423.
EDS is supported by a JSPS Fellowship at RESCEU, and the work of EDS
is supported by Monbusho Grant-in-Aid for JSPS Fellows No.~95209.

\end{document}